\documentclass[draft]{agujournal2019}
\usepackage{url} %
\usepackage{lineno}
\usepackage{soul}
\usepackage{amsmath}

\draftfalse

\newcommand{\degree}{$^\circ$}

\journalname{Geophysical Research Letters}

\begin{document}

\title{Seismic swarms unveil the mechanisms driving shallow slow slip dynamics in the Copiapó ridge, Northern Chile}

\authors{
Jannes Münchmeyer\affil{1,\dagger},
Diego Molina\affil{1,\dagger},
Mathilde Radiguet\affil{1},
David Marsan\affil{1},
Juan-Carlos Baez\affil{2},
Francisco Ortega-Culaciati\affil{3},
Andres Tassara\affil{4},
Marcos Moreno\affil{5},
Anne Socquet\affil{1}
}

\affiliation{1}{Univ. Grenoble Alpes, Univ. Savoie Mont Blanc, CNRS, IRD, Univ. Gustave Eiffel, ISTerre, Grenoble, France.}
\affiliation{2}{National Seismological Center, University of Chile, Santiago, Chile.}
\affiliation{3}{Department of Geophysics, Faculty of Physical and Mathematical Sciences, University of Chile, Santiago, Chile.}
\affiliation{4}{Departamento Ciencias de la Tierra, Facultad de Ciencias Químicas, Universidad de Concepción.}
\affiliation{5}{Department of Structural and Geotechnical Engineering, Pontificia Universidad Católica de Chile, Santiago, Chile.}
\affiliation{\dagger}{These authors contributed equally to this work.}

\correspondingauthor{Jannes Münchmeyer}{munchmej@univ-grenoble-alpes.fr}

\begin{keypoints}
\item We observe a month-long slow slip event ($M_W=6.6$) in Northern Chile, accompanied by dense, migrating seismic swarm activity.
\item The slow slip initiation is likely controlled by migrating fluid overpressure on the segmented interface around a subducted seamount.
\item Historic earthquake swarm activity provides evidence for recurrent slow slip with variable magnitude controlled by the interface structure.
\end{keypoints}

\begin{abstract}
Like earthquakes, slow slip events release elastic energy stored on faults.
Yet, the mechanisms behind slow slip instability and its relationship with seismicity are debated.
Here, we use a seismo-geodetic deployment to document a shallow slow slip event (SSE) in 2023 on the Chile subduction. 
We observe dense, migrating seismic swarms accompanying the SSE, comprised of interface activity and upper plate splay faulting.
Our observations suggest that the slow slip initiation is driven by structurally-confined fluid overpressure in the fluid-rich surroundings of a subducted seamount.
This is consistent with an observed acceleration and expansion of the SSE after a $M_L=5.3$ earthquake likely triggering an increase in interface permeability.
Historical earthquake swarms highlight the persistent structural control and recurrent nature of such slow slip events.
Our observations provide insight into the interactions between slow slip and seismicity, suggesting they are controlled by creep on a fluid-infiltrated fault with fractally distributed asperities.
\end{abstract}

\section*{Plain Language Summary}
Subduction zones continuously accumulate stress from tectonic plate movement.
Part of this stress is released in earthquakes, but another part is released through transient deformations, so-called slow slip events (SSEs).
SSEs lead to the same motion on the plate interface as regular earthquakes, but happen over a much longer timeframe of weeks to months.
SSEs do not cause ground shaking, but they are important to understand the mechanisms leading to major earthquakes.
Here, we report on the first shallow SSE observed in Northern Chile, in the area of the subducted Copiapó ridge.
We combine geodetic observations with a high-resolution earthquake catalog to study the mechanisms of slow slip.
We find dense earthquake swarms, migrating together with the slow slip.
Our findings suggest that the SSE and swarms are both caused by migrating fluid pulses traveling through a fluid-rich plate interface.
The fluid-rich environment results from a subducted seamount on the Copiapó ridge.
The partition of stress release into slow slip and micro-earthquakes can be explained with a segmentation of the plate interface into an aseismic medium with embedded seismic patches.
Our results provide new insight into the mechanisms of SSE initiation and the interplay between slow slip and regular seismicity.

\section{Introduction}

\begin{figure}
\centering
\includegraphics[width=\textwidth]{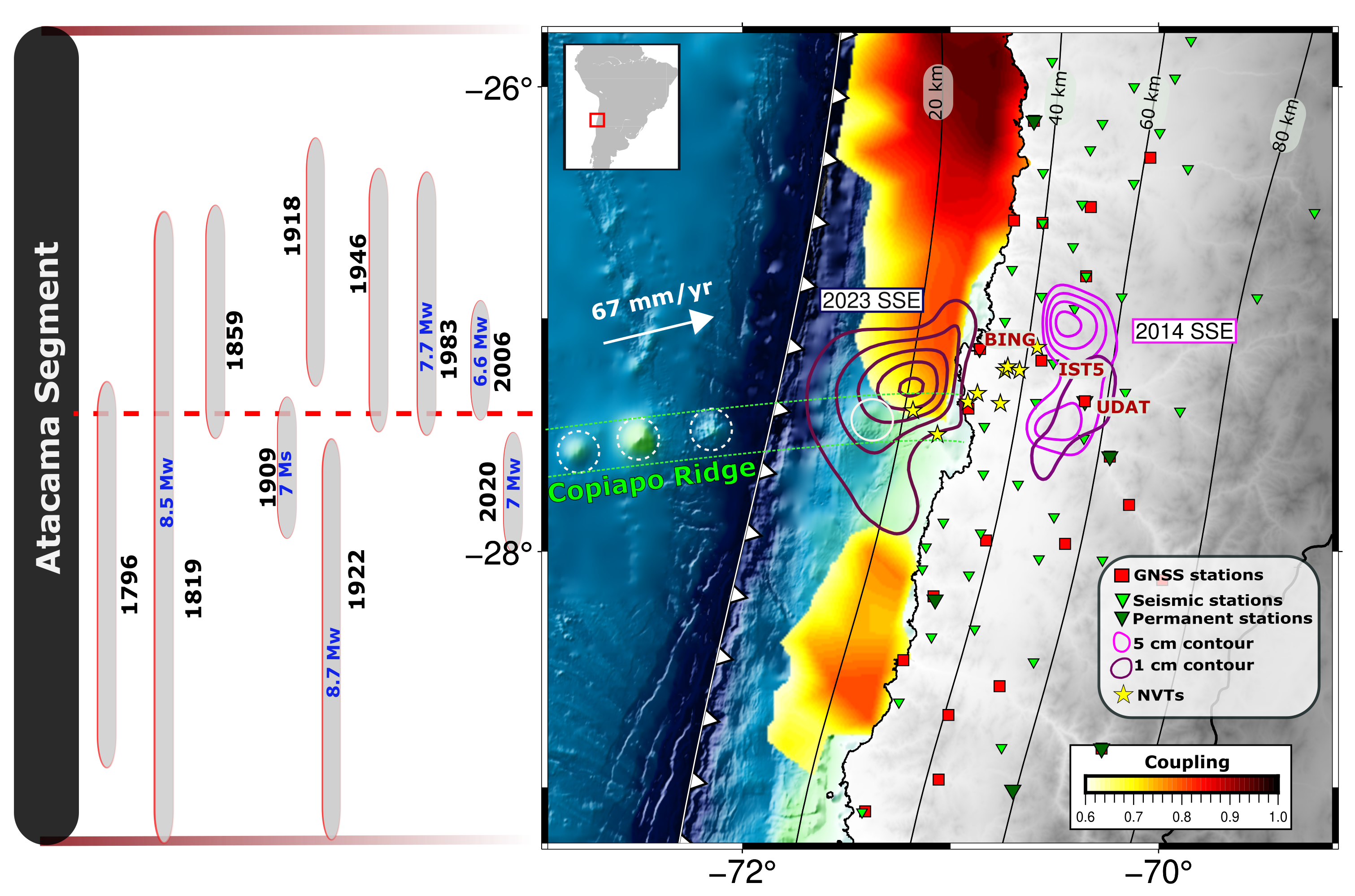}
\caption{Overview of the study region with the Nazca plate subducting below the South American plate. Copiapó ridge highlighted by bathymetry map. Background warm colors show the locking degree \cite{gonzalez-vidalRelationOceanicPlate2023}. Dark green triangles denote permanent seismic stations, light green triangles temporary seismic stations, red squares GNSS stations. Colored contours show the SSE in 2014 (pink, \cite{kleinDeepTransientSlow2018}) and 2023 (maroon, this paper). Yellow stars indicate non-volcanic tremor on 20/09/2019 \cite{pasten-arayaAlongDipSegmentationSlip2022}. On the left, the along strike extent of historic earthquakes is shown \cite{ruizHistoricalRecentLarge2018,vignySearchLostTruth2024}.}
\label{fig:overview}
\end{figure}

Subduction plate interfaces continuously accumulate stress due to tectonic forcing.
This stress is released through earthquakes or slow slip events (SSEs), which last from days to months and typically do not emit strong seismic waves \cite{ideScalingLawSlow2007,takemuraReviewShallowSlow2023,delahayeMicroseismicityNoTremor2009}.
The physical parameters that determine whether a rupture is slow or fast, and the mechanisms driving slow slip instability are still debated \cite{nielsen2017slow,ideScalingLawSlow2007,obaraConnectingSlowEarthquakes2016,kato2021generation,molina-ormazabalDiverseSlipBehaviour2023}.
Furthermore, although slow slip events have been involved in the triggering \cite{radiguetTriggering2014Mw72016,vossSlowSlipEvents2018,uchidaPeriodicSlowSlip2016}, nucleation \cite{obaraConnectingSlowEarthquakes2016,nishikawaRecurringSlowSlip2018,itoh2023largest} or arrest \cite{uchida2013pre,rolandoneAreasProneSlow2018,itoh2023largest} of large earthquakes, how slow slip is related to seismic rupture is still poorly understood.

Several studies suggest that the megathrust interface is segmented into slow and fast slipping regions \cite{kaneko2010towards,molina2021frictional,molina-ormazabalDiverseSlipBehaviour2023} down to small-scale \cite{senoFractalAsperitiesInvasion2003,behrTransientSlowSlip2021,lavierMechanicsCreepSlow2021,yabe2014spatial}.
These segments differ in terms of lithology, geometry, stress heterogeneity, and fluid availability \cite{mangaChangesPermeabilityCaused2012,gosselinSeismicEvidenceMegathrust2020,nakajimaRepeatedDrainageMegathrusts2018,warren-smithEpisodicStressFluid2019}.
While the origin of this segmentation is not fully understood, a commonly proposed mechanism is a structural control by subducted oceanic structures, such as seamounts \cite{singh2011aseismic,barnes2020slow,YanezCuadra2022,gonzalez-vidalRelationOceanicPlate2023,contreras-reyesControlHighOceanic2011}.
To keep the rupture slow,  those segments need to be controlled by mechanisms preventing the acceleration into fast ruptures while maintaining slow ruptures.
Potential mechanisms to initiate and maintain a slow rupture are fluid pressure migrations, frictional heating, or a control through velocity strengthening friction \cite{warren-smithEpisodicStressFluid2019,wangPulselikeRupturesSeismic2023}.
Yet, monitoring these processes and the fine-scale segmentation directly has proven challenging due to insufficient resolution.

Here, we use surface deformation and swarm seismicity associated with an SSE in 2023 in the subduction Copiapó ridge, Northern Chile, to investigate the processes and structures driving the SSE initiation and propagation.
Between November 2020 and Februrary 2024, the Copiapó ridge has been monitored with a dense network of GNSS \cite{Socquet_Metadata_GNSS} and seismic stations.
We use the high-resolution earthquake catalog by \citeA{munchmeyer2024chile_eqs} inferred from the seismic records using deep learning techniques.
The catalog contains very precise event hypocenter estimates: in the SSE area, typical relative errors for events at an interevent distance of 500~m are below 50~m and for events at a distance of 5~km below 100~m \cite{munchmeyer2024chile_eqs} (Figures~S1,~S2).
These observations allow to mitigate the observational gap obstructing our understanding of the mechanisms driving the slow slip instability and its interactions with seismicity.
We are able to monitor with unprecedented detail how the SSE initiated, and what phenomenon triggered the SSE expansion and propagation.
Our study sheds light on the role of structural complexities in trapping pressurized fluids, and on the interplay between slow slip, seismicity, and fluid pressure migrations.

\section{The subducted Copiapó ridge and the 2023 shallow SSE}

\begin{figure}
\centering
\includegraphics[width=\textwidth]{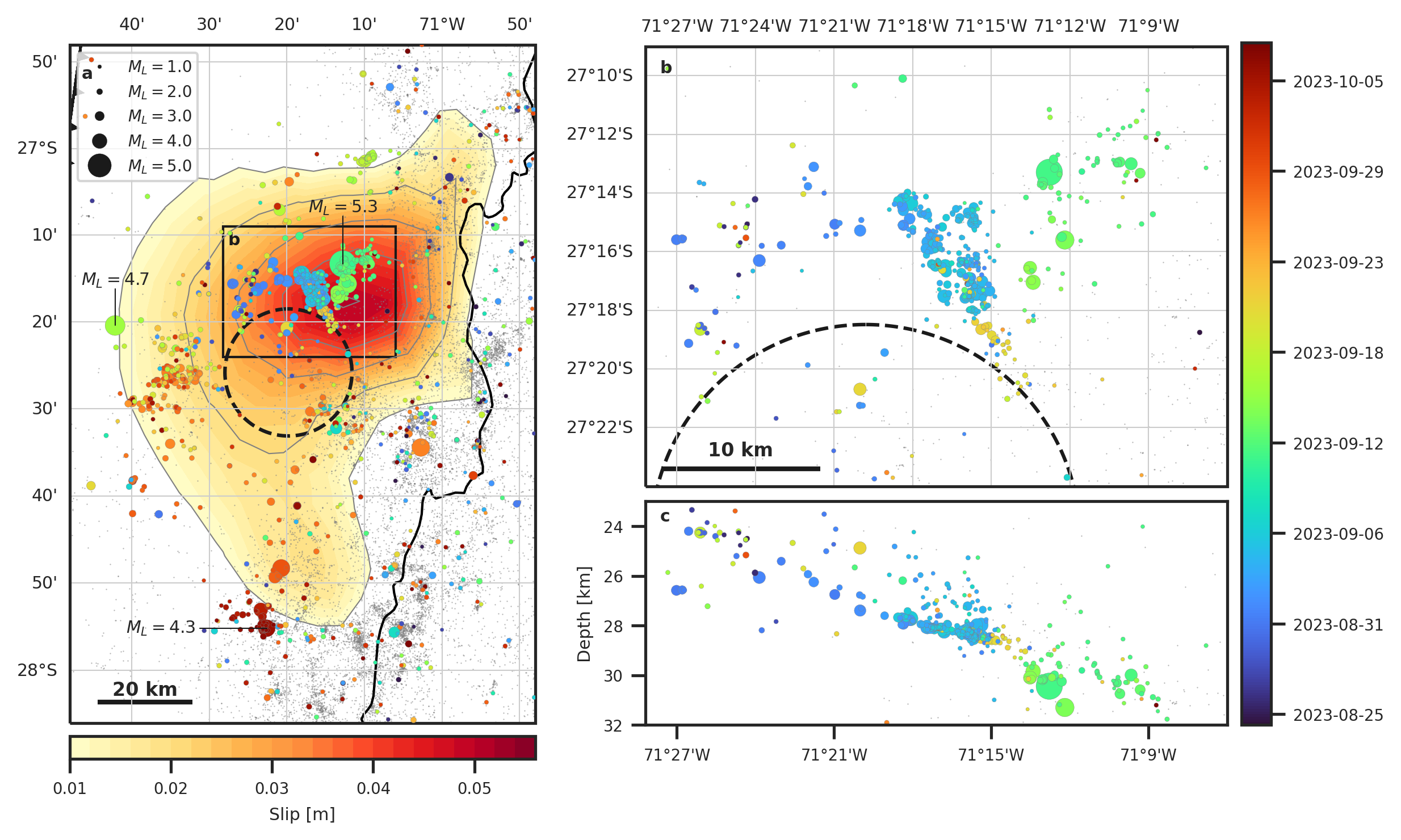}
\caption{\textbf{a} Slip distribution and seismicity during the 2023 slow slip event. The slip distribution is shown in yellow to red with additional grey 1~cm contours. Earthquakes during the time period are colored by time in the sequence and scaled by magnitude. For reference, all seismicity cataloged between 11/2020 and 02/2024 is shown as grey dots. \textbf{b} Zoom in on the onset cluster. \textbf{c} Cross section of the events in b along longitude. The black dashed circle in a and b indicates the approximate position of the subducted seamount inferred from bathymetric features and gravity data.}
\label{fig:sse2023}
\end{figure}

\begin{figure}
\centering
\includegraphics[width=\textwidth]{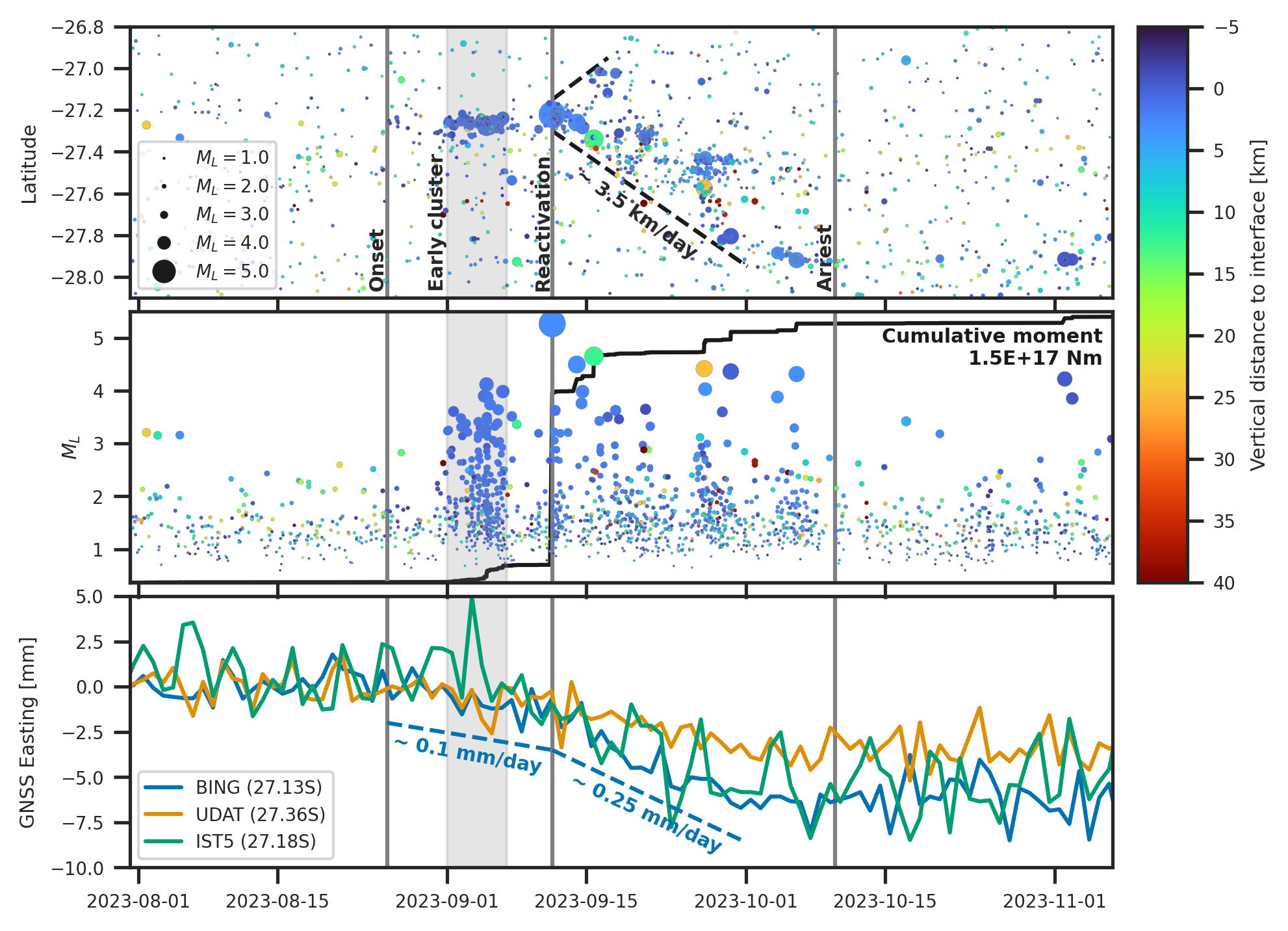}
\caption{Temporal development of latitude distribution, magnitudes, and detrended GNSS series. The top two plots include all events shown in Figure~\ref{fig:sse2023}a. The black line in the magnitude plot shows the total cumulative seismic moment release. GNSS time series have been selected for stations close to the main slip patch of the SSE. We highlight the approximate GNSS velocities at the station BING with blue dashed lines.}
\label{fig:time_latitude}
\end{figure}

At the latitude of Copiapó, an oceanic ridge enters into subduction, as marked by a series of aligned seamounts, corresponding to bathymetric anomalies of 2 to 3~km compared to the surrounding seafloor \cite{maksymowicz2024joint} (Figure~\ref{fig:overview}).
Above the megathrust, a local positive topographic anomaly is most likely the surface expression of a subducted seamount (Figure~\ref{fig:overview}), as already shown in other settings \cite{bangsSlowSlipHikurangi2023}. 
The megathrust around the subducted Copiapó ridge is characterized by a low locking degree  (Figure~\ref{fig:overview}) \cite{gonzalez-vidalRelationOceanicPlate2023} and has repeatedly served as a barrier to seismic ruptures \cite{comteSeismicityStressDistribution2002,ruizHistoricalRecentLarge2018,vignySearchLostTruth2024}.
A large earthquake is pending in this region, as adjacent locked patches have not been ruptured by a large earthquake ($M_w>8$) since at least 1922 \cite{kanamoriNewConstraints19222019,gonzalez-vidalRelationOceanicPlate2023,vignySearchLostTruth2024}.
The region is known to host seismic swarms with reports dating back to 1973 \cite{comteSeismicityStressDistribution2002,ojedaSeismicAseismicSlip2023,marsanEarthquakeSwarmsChilean2023}.
Previous studies have reported deep SSEs (40 to 60~km) \cite{kleinDeepTransientSlow2018,kleinReturnAtacamaDeep2022} with a duration of $\sim1.5$ years and a 5 year return period, and rare occurrences of non-volcanic tremors \cite{pasten-arayaAlongDipSegmentationSlip2022}, but no shallow slow slip had previously been observed.

In September 2023, we recorded a month-long transient motion on 18 GNSS stations between 26.1\degree S and 28.4\degree S with surface displacements up to 8~mm (Figures~S3,~S4).
By inverting the slip on a dislocation in an elastic medium, we infer a shallow SSE in the Copiapó ridge with geodetic $M_w=6.6$ (Figures~\ref{fig:sse2023},~\ref{fig:time_latitude}, Texts~S1-S3).
The associated slip patch extends over $\sim$100~km along strike between 25 and 35~km depths.
The peak slip of $\sim 5$~cm is located on the edge of the inferred subducted seamount.
The 2023 SSE is located updip of the deep SSEs with a band of interface seismicity separating their slip patches \cite{kleinDeepTransientSlow2018,kleinReturnAtacamaDeep2022,munchmeyer2024chile_eqs}.
Although our inversion revealed a small, deep, secondary slip patch with peak slip below 2~cm in the area of the deep 2020 SSE (Figure~S3), it is unclear if this results from actual slip or insufficient resolution.
The SSE starts around September 1st, increases in slip rate in the second half of September, before arresting in early October (Figure~\ref{fig:time_latitude}).
Individual slip inversions for the early ($M_w=6.16$, until 10/09, Figure~S3a) and late periods ($M_w=6.52$, Figure~S3b) of the SSE indicate localized slip around the edge of the subducted seamount in the first phase and a substantially larger slip with bidirectional expansion afterwards.

The SSE is accompanied by seismic swarm activity with complex migration patterns (Figures~\ref{fig:sse2023},~\ref{fig:time_latitude}).
The cumulative seismic moment release is $M_w=5.4$, only about 1.6~\% of the geodetic moment release.
Swarm seismicity starts on August 26 within a well-confined, low-magnitude cluster (total $M_w<3.0$) on the northern edge of a subducted seamount.
On September 1st a vigorous swarm is activated downdip, in the area of highest geodetic slip.
After a five-day quiescence, an interface earthquake $\sim5$~km further downdip ($M_L=5.3$) marks the acceleration of the geodetic displacement.
It also marks the onset of a large-scale, bilateral migration of the seismicity at a rate of $\sim3.5$~km/day (Figure~\ref{fig:time_latitude}), a typical migration rate of month-long SSEs \cite{danrePrevalenceAseismicSlip2022,danreParallelDynamicsSlow2024}.
The short northward migration terminates after 20~km with three magnitude 3.5 earthquakes within 2 days.
The southward migration terminates around 27.9\degree S with the last large event ($M_L=4.3$) on October 6th.
The migration activates several seismicity clusters surrounding the seamount, among which the largest occurs updip of the inferred slip area, including one $M_L=4.7$ event.
Within the area of the seamount, only few isolated events occur (Figure~\ref{fig:sse2023}), yet these fit the surrounding seismic migration patterns.

The seismo-geodetic joint monitoring suggests two different phases during the SSE (Figure~\ref{fig:time_latitude}).
In the initiation phase, up to the $M_L=5.3$ event and the geodetic acceleration, seismicity and slip colocate in a small area.
In the second phase, the slow slip and the seismicity are migrating over larger distances at a constant velocity, with the seismic activity primarily surrounding the main slip patch.
Such two-phase SSE behaviour has previously been observed, for example, in Cascadia \cite{gombergReconsideringEarthquakeScaling2016,itohImagingEvolutionCascadia2022}.
To understand the SSE's initiation mechanisms, we now focus on the seismic activity before the expansion of geodetic slip.

\section{Seismic swarms outlining the SSE initiation}

The full initiation sequence of the SSE occurs along the northern flank of the subducted seamount (Figure~\ref{fig:sse2023} b,c).
It commences with an initial weak burst of seismicity on August 26th (blue in Figure~S5).
From September 1st to 6th a second cluster is activated downdip (cyan), confined to a narrow area of approximately 10~km by 3~km.
Within this swarm, a double migration occurs: the main activity migrates southwards with a slowly decaying rate between 2 and 5~km/day.
It is intersperse with bursts migrating with velocities above 15~km/day (Figure~S6).
Such double migrations and interactions between slow slip and seismicity are indicative of pressure pulses triggering asperities on creeping faults infiltrated by fluid \cite{dublanchetDualSeismicMigration2021}.

This second swarm is clearly delineated downdip by a gap in the seismicity.
Activity beyond this gap only occurs after a 5 day quiescence and initiates with an $M_L=5.3$ earthquake on September 11th.
Within the first two days, the aftershocks of this event (green) expand within an area of around 75~km$^2$ with an expansion rate typical for afterslip \cite{perfettiniModelAftershockMigration2018} (Figure~S7).
The three onset clusters show qualitatively different magnitude distributions (Figure~S5).
The third cluster, containing the largest event, show approximately Gutenberg-Richter distributions.
In contrast, the second cluster, with by far the highest event count, shows a clear scale break around $M_L=3.2$.
This scale break suggests spatially confined ruptures, for example, due to limited size of the seismic asperities \cite{senoFractalAsperitiesInvasion2003,behrTransientSlowSlip2021}.

Within this early sequence, almost all events locate on the narrow plate interface (Figure~S6).
In addition, we are able to resolve earthquakes on splay fractures trending westward in the upper plate.
The seismic activity shows that the splay fractures extend over several kilometers.
Timing of upper plate seismicity coincides with the activation of interface seismicity and is most active above the second cluster (cyan in Figure~S5).
This activity of splay fractures might be indicating the drainage of fluid overpressure from the interface \cite{chesleyFluidrichSubductingTopography2021,perez-silvaCharacteristicsSlowSlip2023}.

\section{Revisiting historic swarm activity}

\begin{figure}[ht!]
\centering
\includegraphics[width=0.75\textwidth]{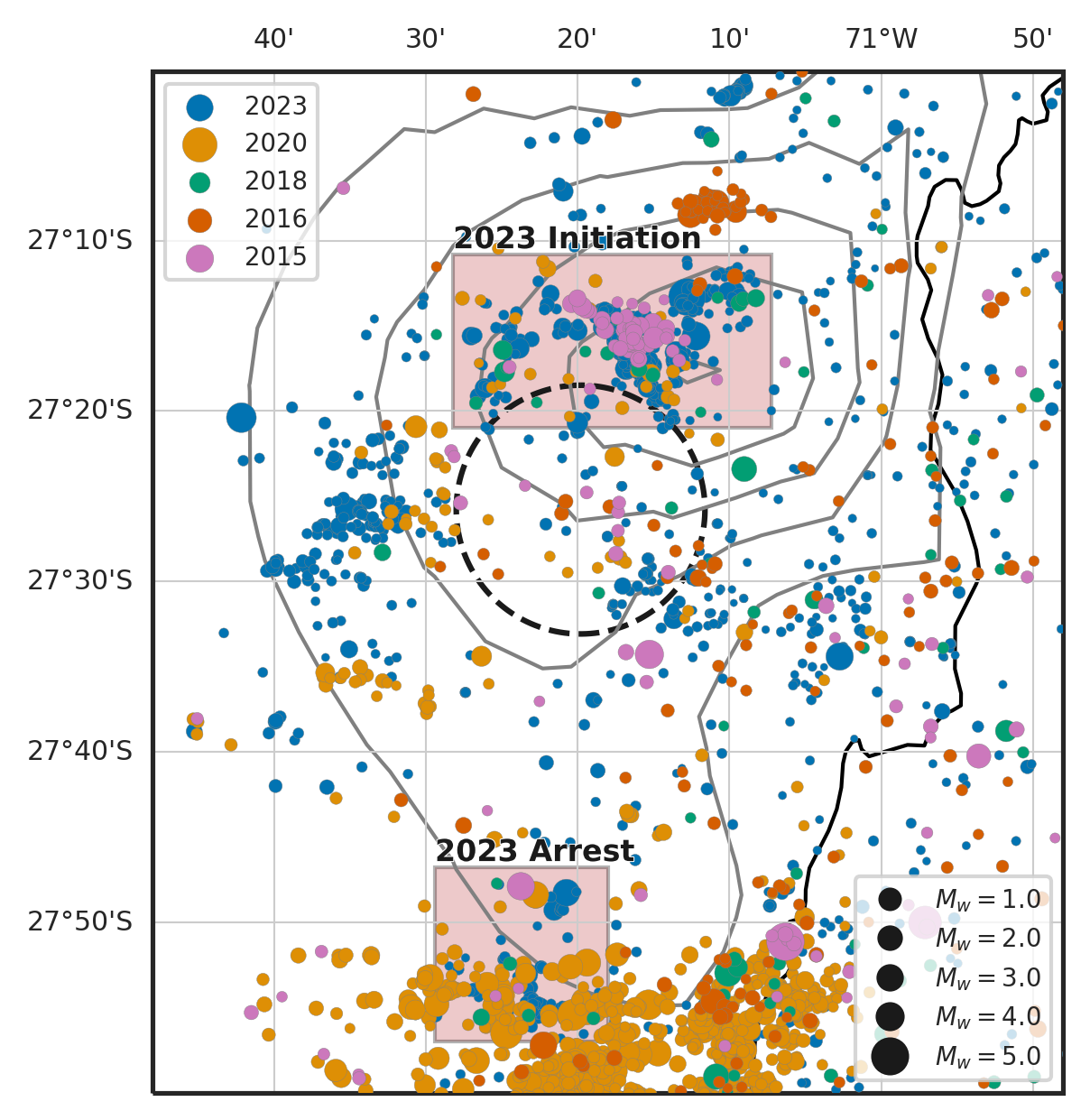}
\caption{Zoom in on the central part of the 2015, 2016, 2018, 2020 and 2023 swarms using catalogs from \citeA{munchmeyer2024chile_eqs}. The black circle highlights the approximate location of the subducted seamount. Grey lines show the 1~cm slip contours of the 2023 SSE. All swarms have been located with the same 3D seismic velocity model. The 2015 to 2020 swarms use similar station coverage, consisting of permanent seismic networks, while for the 2023 swarm a collection of much denser temporary networks was used. Therefore, catalog completeness and location accuracy are best for the 2023 swarm. The 2015 to 2020 swarms have been jointly relocated using GrowClust3D, the 2023 model individually.}
\label{fig:historic_swarms_v2}
\end{figure}

In order to investigate whether such events are recurrent and if any repetitive pattern is found, we compare the swarm activity in 2023 to nine historic swarms around the Copiapó ridge.
These swarms encompass newly detected sequences and known swarms using four seismic catalogs (Figures~\ref{fig:historic_swarms_v2},~S6) \cite{comteSeismicityStressDistribution2002,ojedaSeismicAseismicSlip2023,marsanEarthquakeSwarmsChilean2023}.
Due to limited detection capabilities, especially prior to 2014, we likely miss additional swarms in between the early swarms.
The five swarms with sufficient data (2015, 2016, 2018, 2020, 2023) have been located using a common 3D velocity model.
These swarms consistently activate the same interface patches, in particular, in the initiation phase around the seamount and in the arrest region of the 2023 SSE (Figure~\ref{fig:historic_swarms_v2}).
The active patches match at least down to the kilometer scale, similar to the location uncertainties of the longer-term catalog used for the swarms from 2015 to 2020.
This suggests that the location of the swarm seismicity is controlled by persistent interface structures that are repeatedly activated.
This observation is consistent with the previous observation that the 2023 initial swarm is structurally confined.

While geodetic records are insufficient to detect potential SSEs, several indicators point at slow slip associated with the swarm activity.
Most prominently, all swarms exhibit migration patterns (Figure~S8).
This is particularly evident for the 1973 swarm (northward migration) and the 2006 swarm (bilateral).
Furthermore, for the 2006 sequence, one continuous GNSS station running during the swarm detected seaward motion of the upper plate, consistent with a shallow SSE \cite{ojedaSeismicAseismicSlip2023}.
For the 1973 event, moment tensor solutions show that the events features thrust mechanisms, indicating that the seismicity is mostly located on the subduction interface \cite{comteSeismicityStressDistribution2002}.
Notably, the region shows almost no seismic activity in between the seismic swarms.
In combination with the observed migration patterns, this suggests that swarm seismicity is caused by outside drivers, such as recurrent SSEs.

The detected swarms vary in seismic moment by almost three orders of magnitude.
As geodetic moment typically scales with seismic moment, this suggest that the region possibly hosts repeated SSEs with highly varying moment release \cite{nishikawaRecurringSlowSlip2018,passarelliSourceScalingSeismic2021,wangPulselikeRupturesSeismic2023}.
Given this observations, it seems possible that the low locking in this area is not the result of continuous creep, but that the slip pattern is highly intermittent, with episods of large slip detectable as SSEs with accompanying seismic swarms.
Such intermittence on short time-scales is not resolvable in on-shore geodetic data, and will require offshore deployments to verify.

\section{The structure of the subduction interface}

\begin{figure}
\centering
\includegraphics[width=\textwidth]{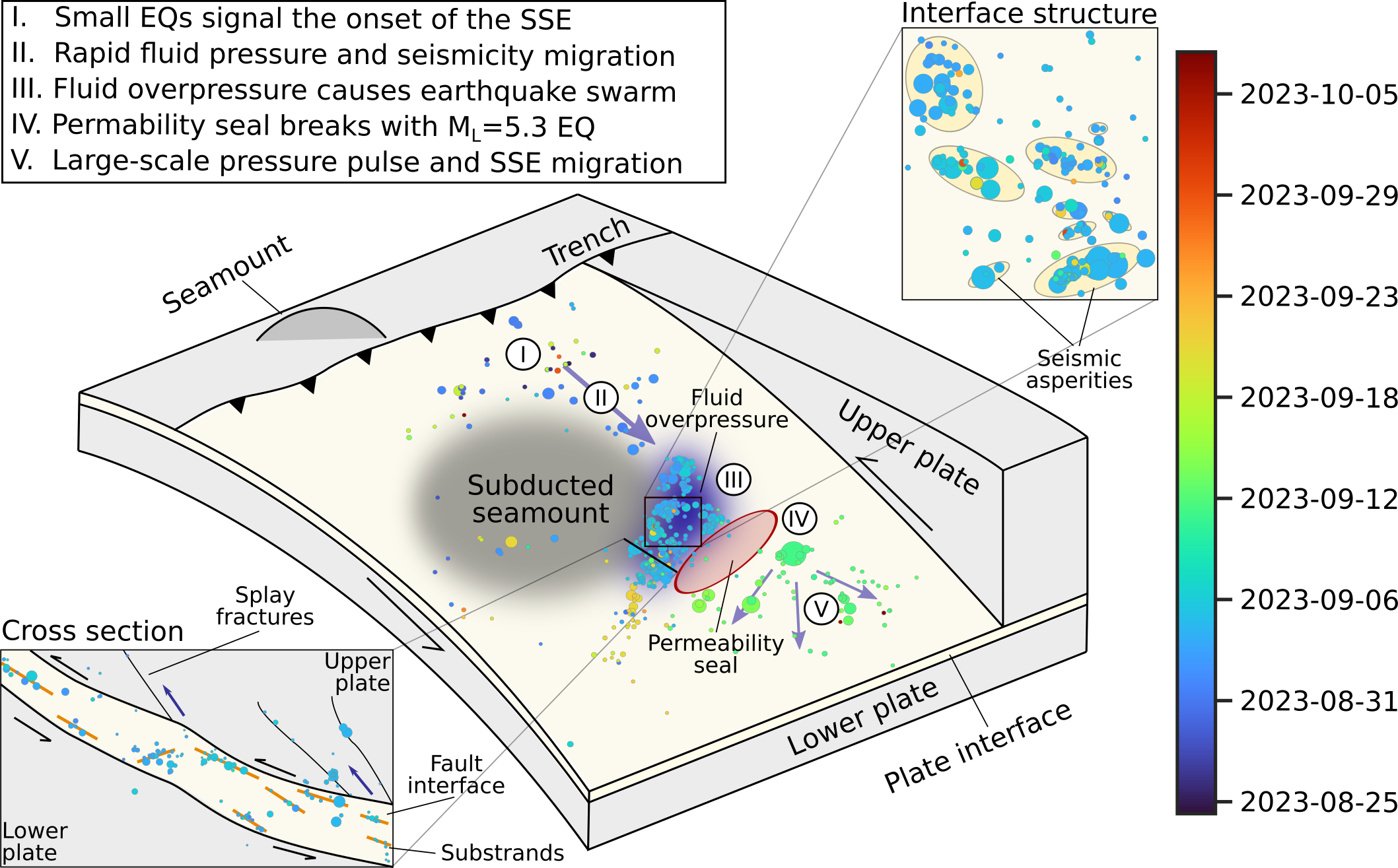}
\caption{
Schematic overview of the SSE initiation sequence and its drivers. Each earthquake is represented by a dot colored according to the time within the sequence. Blue arrows show fluid pressure migrations. The insets show a cross section of the fault interface (bottom left) and the interface structure in bird's eye perspective (top right).}
\label{fig:scheme}
\end{figure}

Based on our observations, we infer a schematic model of the SSE area (Figure~\ref{fig:scheme}).
The high resolution of the earthquake catalog shows that the subduction interface is only seismically active on small patches (Figure~\ref{fig:scheme} inset), each of them extending up to 50~m vertically and up to 2~km horizontally.
We observe rapid migrations within these patches.
The individual patches are disconnected from each other and distributed within a thin, 3D fault interface volume of less than 250~m thickness, overall forming the subduction interface on a macroscopic scale (Figure~S6 right).
This is consistent with estimates of the interface thickness in other subduction zones \cite{chalumeauSeismologicalEvidenceMultifault2024}.
The patches show a fractal behaviour from a length scale of 5~km down to the location uncertainty around 40~m, meaning that the seismic patches cluster and are themselves subdivided into smaller seismic patches (Figure~\ref{fig:scheme} inset, Figure~S1).
Our comparison to historic swarms shows that this segmentation into seismic and aseismic patches is stable at least on the decade scale.
As the total seismic moment is insufficient to accommodate the geodetic moment, the seismic patches need to be embedded within a slowly deforming matrix \cite{senoFractalAsperitiesInvasion2003,perfettiniModelAftershockMigration2018,behrTransientSlowSlip2021,lavierMechanicsCreepSlow2021}.
This observation is consistent with geological observations of fossil subduction interfaces, a shear zone with frictional lenses embedded in a viscous and fluid-rich matrix \cite{behrTransientSlowSlip2021}.

To explain the episodic, swarm-type seismic behaviour and the transient slow deformation, we suggest a central role of pressure pulses migrating on a fluid-rich interface  (Figure~\ref{fig:scheme} I-III).
Such recurrent fluid pressure increases are evidenced by swarm seismicity, the characteristic double-migration patterns (Figure~S6) \cite{dublanchetDualSeismicMigration2021}, the activation of splay fractures for fluid pressure drainage (Figure~\ref{fig:scheme} cross-section)\cite{fasolaEarthquakeSwarmsSlow2019,chesleyFluidrichSubductingTopography2021}, and the typical migration velocities (Figure~\ref{fig:time_latitude}) \cite{danrePrevalenceAseismicSlip2022}.
Furthermore, the seismic quiescence between September 6th and 11th suggests the existence of a structural permeability seal (Figure~\ref{fig:scheme} IV), \cite{mangaChangesPermeabilityCaused2012,gosselinSeismicEvidenceMegathrust2020}.
Before its rupture, the overpressured fluids, structurally confined by the seal, lower the effective normal stress and thereby trigger the most active seismicity swarm in the sequence.
This fluid overpressure infiltrating the size-constrained asperities explains the scale-break observed in the magnitude distribution of the onset cluster (Figure~S5).
Once shear stress on the asperities is released, the cluster goes dormant, even though fluid pressure stays high.
During this cluster and the subsequent quiescence, the aseismic slip remains confined to the initiation area.
Only once the permeability seal is broken, as marked by the $M_L=5.3$ event, the fluid pressure pulse can migrate further along the interface, triggering the expansion of slow slip on the interface together with the large-scale seismic migration (Figure~\ref{fig:scheme} V).
During that propagation stage, the seismicity is produced by the rupture of the seismic asperities as a response to their transient loading by the propagation of the slow slip front.
This is again compliant with a model of seismic asperities embedded in a slowly creeping matrix \cite{yabe2014spatial}.
The propagation of the slow slip is likely arrested by stronger seismic asperities \cite{yabe2014spatial}, as shown by the peripheral location of the seismic swarms with the largest magnitudes compared to the area slipping slowly and inferred by GPS (Figure \ref{fig:sse2023}).

The initiation of the slow slip event, as imaged by the seismic swarm activity, occurs along the northern flank of a subducted seamount (Figure~\ref{fig:sse2023}).
Abundant evidence suggests that subducted ridges stay intact after subduction, leading to zones of low coupling in its wake, and enhance increased normal stress on its leading edge and weakening of the surrounding medium through fracturing \cite{singh2011aseismic,barnes2020slow,contreras-reyesControlHighOceanic2011,sunMechanicalHydrologicalEffects2020}.
Furthermore, seamount flanks are typically observed to feature overpressurized sediments, caused by the subducting topography and have been shown to supply fluids to the interface \cite{shaddoxSubductedSeamountDiverts2019,Gase2024subducting,bangsSlowSlipHikurangi2023,chesleyFluidrichSubductingTopography2021,sunMechanicalHydrologicalEffects2020}.
We hypothesise that this high fluid pressure on the interface enables the migrations of pressure pulses as described above \cite{cruz-atienzaRapidTremorMigration2018}.
Further evidence for a fluid-rich environment is provided by tomographic studies, showing high vp/vs ratios in this region \cite{pasten-arayaAlongDipSegmentationSlip2022}.
Notably, while in our observation the highest activity occurs on the northern flank of the seamount, we see further swarms on its western and south-eastern edges, suggesting similar interactions of interface segmentation and fluid overpressure.
Combining these aspects, the seamount together with the inherited oceanic fractures likely controls the interface segmentation and creates a fluid-rich environment, enabling seismic swarm activity and slow deformation.

A few peculiarities should be noted regarding the swarm behaviour and its relation to fluid pressure: the downdip migration, the roughly constant migration velocity, and the high value of the migration velocity.
All three aspect would be unexpected for fluid diffusion, but are fully compatible with pressure pulses on a fluid-rich interface during slow slip \cite{cruz-atienzaRapidTremorMigration2018,perez-silvaCharacteristicsSlowSlip2023,fargeEpisodicityMigrationLow2021,fargeAlongStrikeSegmentationSeismic2023}.
This is caused by transient changes in permeability, due to changes in effective pressure and dilatancy from the slow slip, and by anisotropy of the permeability \cite{kawanoPermeabilityAnisotropySerpentinite2011}.
We note that this is not the only interaction mode of slow slip with seismic swarms.
Alternative modes are intraplate swarms in the subducting plate, related to the fluid supply to the interface \cite{nishikawaEarthquakeSwarmDetection2021}, and swarms on upper plate faults caused by fluid drainage \cite{fasolaEarthquakeSwarmsSlow2019}.

The micro-structures and pressure pulse migrations define the macro-structure of the subduction megathrust and its seismic potential.
Differential stresses built up by repeated SSEs increase the likelihood of triggering large earthquakes in adjacent segments \cite{radiguetTriggering2014Mw72016,obaraConnectingSlowEarthquakes2016}.
At the same time, the predominantly aseismic interface acts as an energy sink for seismic rupture propagation \cite{kaneko2010towards,molina2021frictional}, explaining the barrier behaviour of this region \cite{ruizHistoricalRecentLarge2018,vignySearchLostTruth2024}.
The seismic patches encircling the seamount indicate that, although the Copiapó ridge generally acts as a barrier, it might still break during a megathrust rupture with sufficient along-dip extent \cite{rice1996slip,YanezCuadra2022}.

\section{Conclusion}

Our findings combine dense seismo-geodetic observations to study the 2023 Copiapó SSE and its seismic swarm activities.
Using a high-resolution catalogue, we gain a detailed view of the mechanisms of shallow slow slip and the subduction interface structure.
Our observations suggests that the interplay of a fine-scale interface segmentation with migrating fluid pressure pulses controls both slow slip and the associated swarm seismicity.
Notably, our study describes the initiation phase of a slow slip event with unprecedented detail.
The slow slip instability initiates with a structurally controlled seismic swarm driven by pressurized fluids, that eventually triggers the acceleration and propagation of the slow slip after the rupture of a permeability seal.
The link between slow slip and seismicity significantly differs between the initiation and the propagation phases of the SSE, indicating that the driving mechanisms evolve from the slow slip initiation to its propagation.
Our results also provide a detailed perspective on the functioning of a seismic barrier, and thereby the understanding of fast and slow megathrust earthquakes.

\section*{Open Research Section}
The seismicity catalog and geodetic data will be made public upon acceptance of the manuscript.

\acknowledgments
We thank everyone involved in the deployment of the seismic and geodetic networks and the data management. 
The computations presented in this paper were performed using the GRICAD infrastructure (\url{https://gricad.univ-grenoble-alpes.fr}), which is supported by Grenoble research communities.
This work has been partially funded by the European Union under the ERC CoG 865963 DEEP-trigger and the grant agreement n°101104996 (“DECODE”). Views and opinions expressed are however those of the authors only and do not necessarily reflect those of the European Union or REA. Neither the European Union nor the granting authority can be held responsible for them.
D.M.'s salary was covered by a grant from CNES.
J-C.B., F.O-C. and M.M. acknowledge funding from ANID PIA-ACT192169 Anillo PRECURSOR project. F.O-C. acknowledge support from FONDECYT 1231684 ANID project. M.M. acknowledge support from FONDECYT 1221507 ANID project. J-C-B. acknowledge support from FONDECYT 1240501 ANID project . Stations from the 9C network were funded by CNRS-Tellus "COP2020: SlowSlip Trigger" and ANR-S5 (ANR-19-CE31-0003) projects.

\bibliography{mybibfile}

\appendix

\clearpage
\appendix

\section*{Supplementary material}
\renewcommand\thefigure{S\arabic{figure}}
\renewcommand\thetable{S\arabic{table}}
\setcounter{figure}{0}
\setcounter{table}{0}

\noindent\textbf{Text S1 - GNSS data availability and processing}

In this work we benefit from Chilean and international efforts to increase the density of the geodetic network monitoring the Atacama Segment. The large number of GPS stations gives high spatial resolution to capture the diversity of tectonic processes taking place. In our study, we analyse 5 permanent stations provided by the CSN and 12 temporary stations deployed by external projects. Furthermore, an additional 4 Chilean and 9 Argentinean stations are used to remove potential common noise from our data. The time window used in our analysis spans roughly 3 years, which is sufficient to recover the current interseismic loading rate. The GNSS data have been processed using GipsyX \cite{Bertiger2020GipsyX} software following a strategy described in \cite{lovery2024heterogeneous}. Time series in IGS20 reference frame used in this manuscript are available at \url{https://doi.osug.fr/public/GNSS_products/GNSS.products.SouthAmerica_GIPSYX.daily.html}.  

\noindent\textbf{Text S2 - GNSS data analysis}

\begin{figure}[ht!]
\centering
\includegraphics[width=\textwidth]{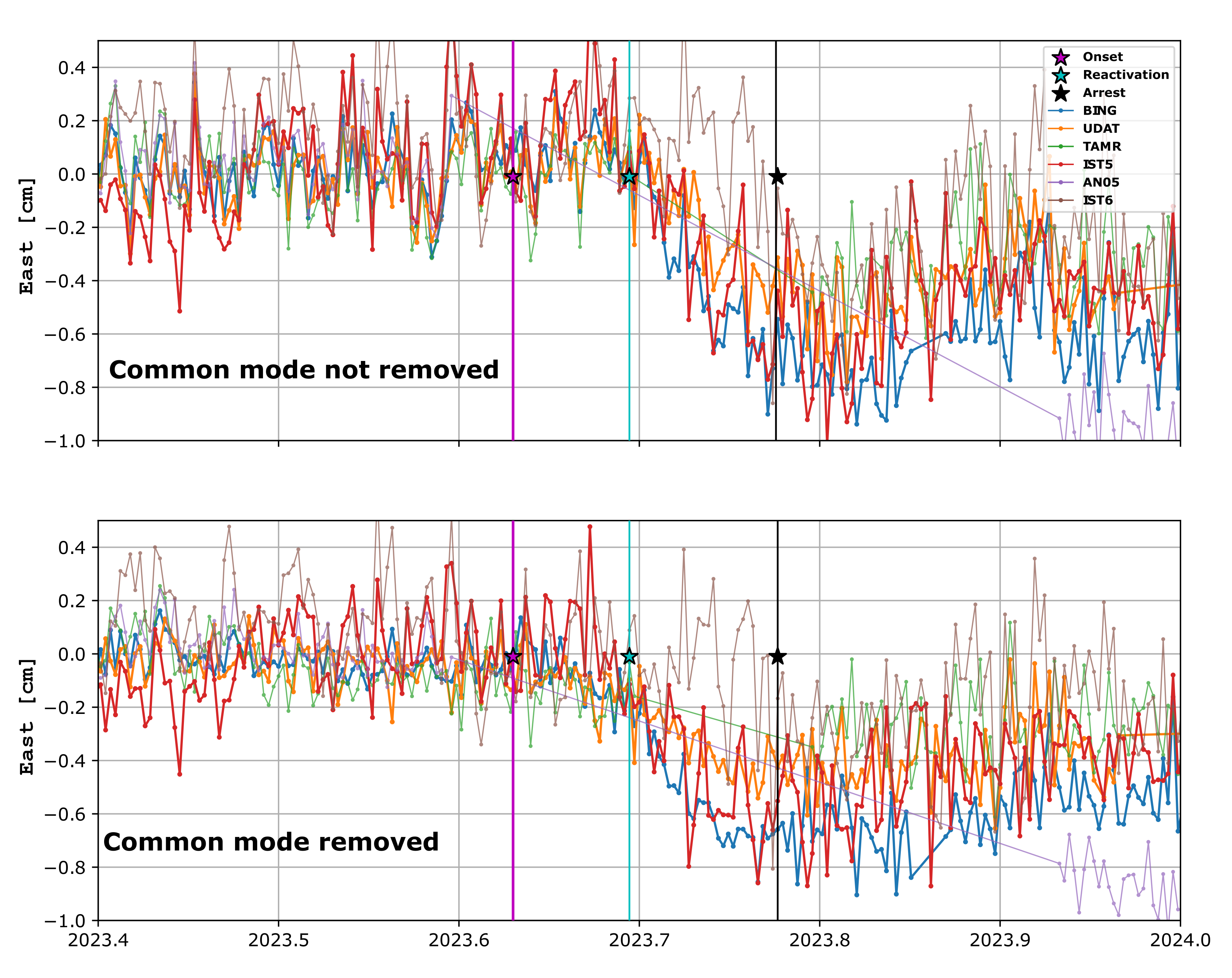}
\caption{Comparison of the GNSS displacement with and without common mode removal. Top: Detrended GNSS time series before removing the common mode. Bottom: Detrended GNSS time series after removing the common mode.}
\label{fig:impact_common_mode}
\end{figure}

\begin{figure}[ht!]
\centering
\includegraphics[width=\textwidth]{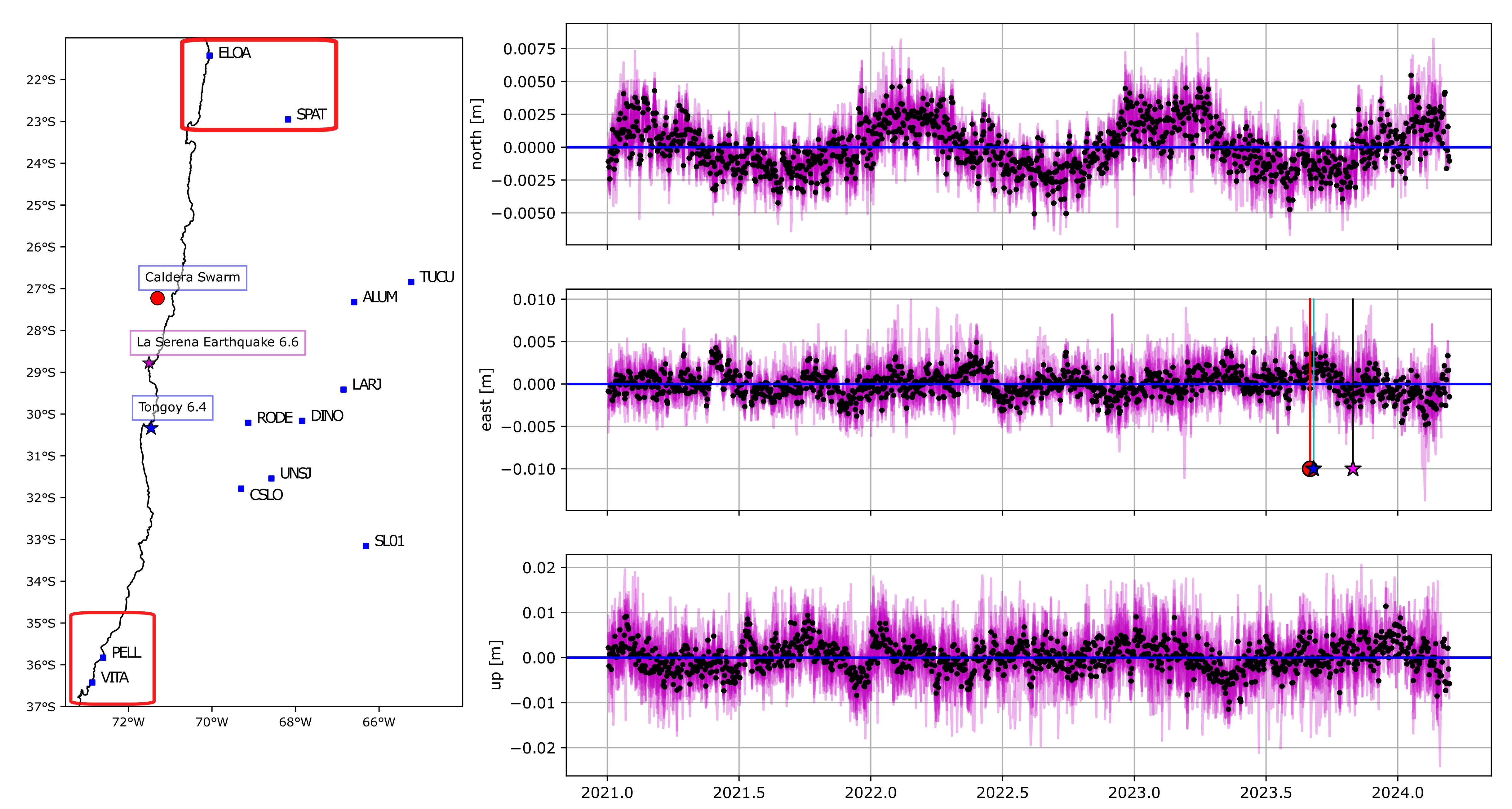}
\caption{Common mode calculated from distant GNSS stations. Left: Map view GNSS stations used to compute the common mode. Red rectangles indicate the 4 Chilean GPS stations. Right: Stacked GNSS time series in pink color. Black dots depict the averaged common mode.}
\label{fig:common_mode}
\end{figure}

Extracting low amplitude transient deformation from GNSS time series can be challenging, specially in areas highly affected by interseismic deformation, seasonal deformation and recurrent earthquakes. In this context, one first task to be done is detecting jumps in the data related to seismic events or antenna change. To extract these steps, we fit a trajectory model in the position time series, using the ITSA package (\url{https://gricad-gitlab.univ-grenoble-alpes.fr/isterre-cycle/itsa}) \cite{marill2021fourteen}. Once these contributions (jumps) have been removed from the time series, the common mode should be corrected, i.e., the temporal evolution that is common to several GNSS stations mostly due to reference frame errors (uncertainties in orbits or earth orientation parameters) or to large scale hydrological signals. To compute the common mode, we stack the detrended position time series at 13 stations located at distances above 200 km from the study area (9 stations belonging to the Argentinean Geodesy Service and 4 Chilean stations (Figure~\ref{fig:common_mode}) in order to ensure that none of them are affected by the targeted tectonic phenomena. The common mode signal is characterized by seasonal variations on the north component, and by high frequency noise in the east and vertical components (Figure~\ref{fig:common_mode}). The common mode is then substracted from all GNSS position time series, decreasing the noise and highlighting more clearly the different transient signals (Figure~\ref{fig:impact_common_mode}). 

In order to highlight the transient signal, the denoised position time series are then detrended over the 2 years preceding the SSE (09/2021 - 09/2023). In order to not be affected by the 2020-2021 deep SSE that occurred in the same area, we did not detrend using a longer time period. We obtain a new set of detrended GNSS time series with roughly zero displacement until middle August 2023 (Figure~\ref{fig:impact_common_mode}). Late August, several stations around the Copiapo Ridge show a slow westward (seaward) displacement. The signal is most evident in the east component, with maximum amplitude in BING and UDAT stations (Figure~\ref{fig:impact_common_mode}). The signal ends in early October, as shown by a flattening of the position time series. 

\begin{figure}[ht!]
\centering
\includegraphics[width=\textwidth]{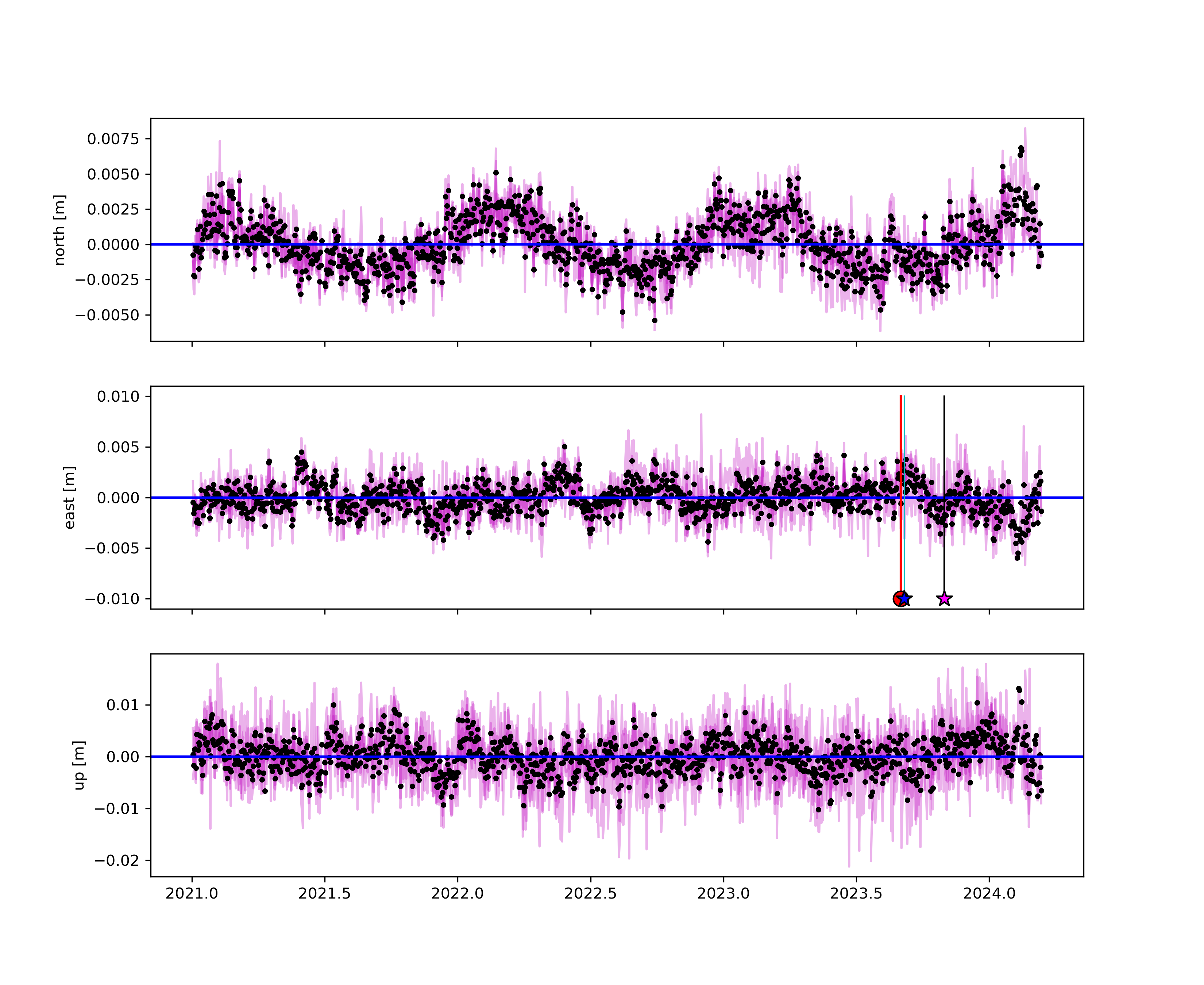}
\caption{Common mode calculated from Chilean GNSS stations (Red rectangles Figure~\ref{fig:common_mode}, Roughly 500 km from targeted area). Stacked GNSS time series in pink color. Black dots depict the averaged common mode.}
\label{fig:common_mode_chilean_stations}
\end{figure}

We performed the same analysis without removing the common mode to evaluate how much the common mode analysis impacts the signal obtained for the SSE. Surprisingly, the signal displays a longer duration and increased amplitude over almost all the stations (Figure~\ref{fig:impact_common_mode}), most evident on the east component. We found that the common mode related to the east component depicts a strong deflection during the same period of time as the SSE (September-October 2023). To ensure that the eastern Argentinean station do not contain this transient, we performed an additional common mode estimation with only those station far north and south. Our findings show that the same signature and pattern result in this new analysis (Figure~\ref{fig:common_mode_chilean_stations}). Therefore not removing the common mode pattern, would have provoked an overestimation of the SSE.

To extract the static offset displacements, we fit our data with an arctangent function. The function is similar to the one proposed by \cite{larson2004crustal}, parametrized by the duration $t_d$, the start time $t_i$ of the SSE, and two amplitude coefficients $a$ and $b$.
The displacement at station $n$ on component $j$ is modelled as:
\begin{align*}  
SSE_{nj}(t) = a_{nj} \arctan(2\pi\frac{t-t_i}{t_d}) + b_{nj}
\end{align*}
The times $t_d$ and $t_i$ are fit jointly at all stations, the constants $a$ and $b$ independently.

\noindent\textbf{Text S3 - Slip inversions}

\begin{figure}[ht!]
\centering
\includegraphics[width=0.5\textwidth]{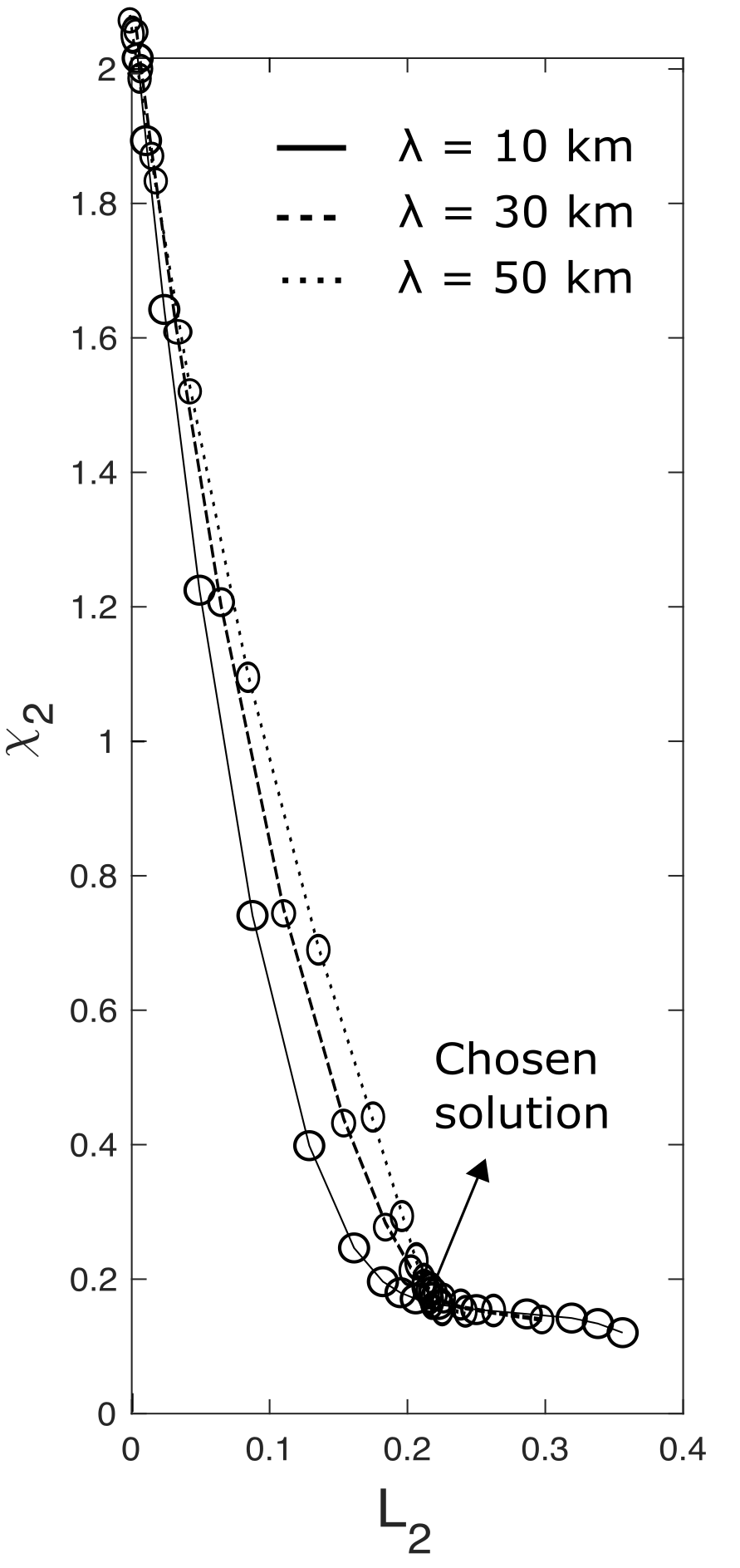}
\caption{L curves criteria used in this study to select our preferred slip model.}
\label{fig:L-Curves}
\end{figure}

To estimate the slip on the plate interface, we perform a linear least square inversion described in \cite{tarantola2005inverse}. This strategy has been adapted following \cite{radiguetTriggering2014Mw72016}. Assuming an elastic half space and a realistic slab geometry (Slab2.0 \cite{hayesSlab2ComprehensiveSubduction2018a}), we calculate the Green's functions \cite{okada1992internal}. We discretize the fault into regular triangular patches of 10~km side length. We impose the slip rake in the convergence direction. We invert the surface displacements computed from the arc tangent function mentioned before. For estimating the seismic moment, we assume a shear modulus of 30~GPa.
To avoid instabilities inherent to the slip inversion problem\cite{OrtegaCulaciati2021}, we set a smoothing prior on slip.  For such purpose, we define the \textit{a priori} model covariance matrix following \cite{radiguetTriggering2014Mw72016}. The covariance matrix depends on two damping constants, $\sigma_{mo}$ and $\lambda$. The value of $\sigma_{mo}$ represent the \textit{a priori} standard deviation on fault slip. The value of $\lambda$ represents a correlation function between neighboring fault patches, aiming to solve for a spatially smooth slip distribution. We imposed $\lambda = 10$ km which results in a more rough solution, with low smoothing. To select an optimal regularization, we select $\lambda$ considering the data fit $\chi^2$ and the $L_2$ norm of the regularized solution. Figure~\ref{fig:L-Curves} shows the L-curves \cite{hansen1993use} for our slip inversion.

\begin{figure}[ht!]
\centering
\includegraphics[width=0.7\textwidth]{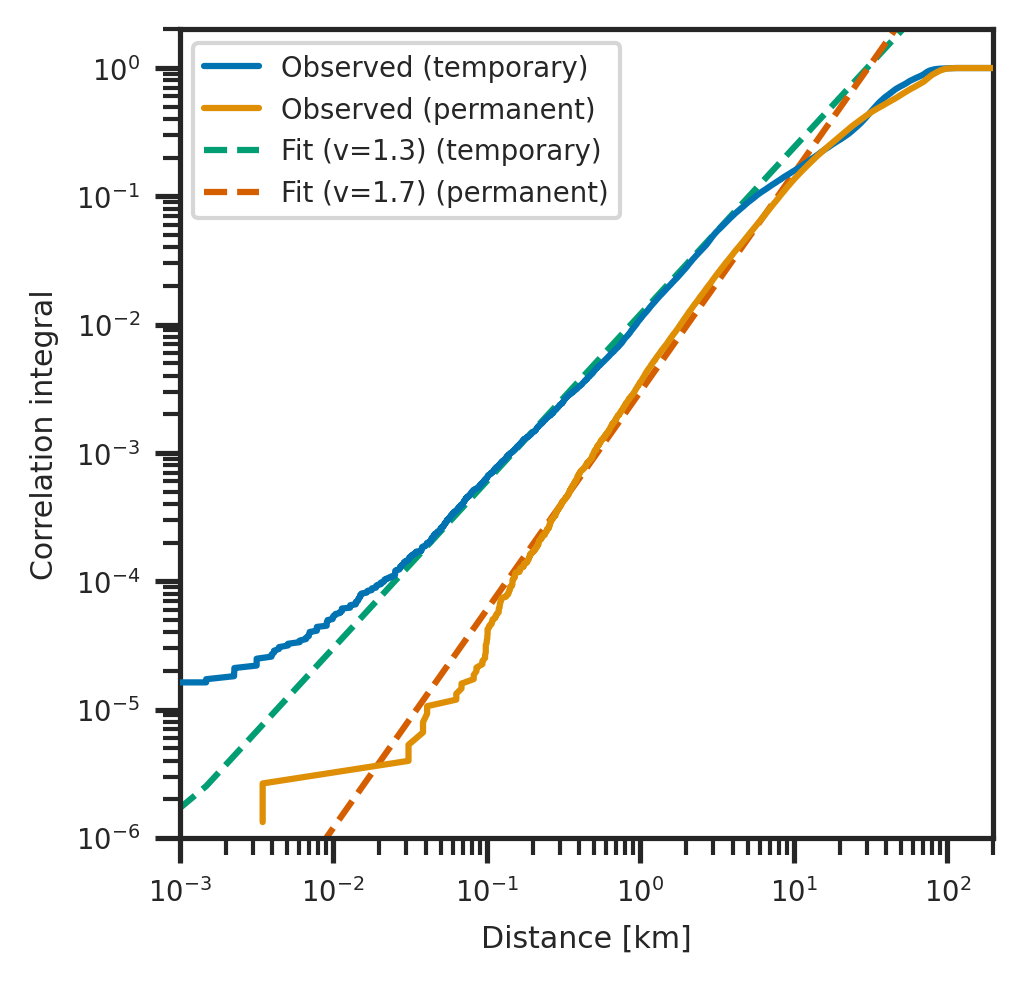}
\caption{Correlation integral over distance range for the temporary and permanent catalogs. We calculate the correlation integral for all events in the boxes in Figure~\ref{fig:historic_swarms}, i.e., events that occur during swarm episodes and within area of the 2023 SSE. The dashed lines shows a log-log fit to determine the fractal dimension. We suspect that the higher fractal dimension for the permanent catalog is caused by the different seismicity pattern in the aftershock region of the 2020 Coquimbo earthquakes, as these events make up a large fraction of the total events in the permanent catalog in this region.}
\label{fig:fractal_dimension}
\end{figure}

\begin{figure}[ht!]
\centering
\includegraphics[width=0.9\textwidth]{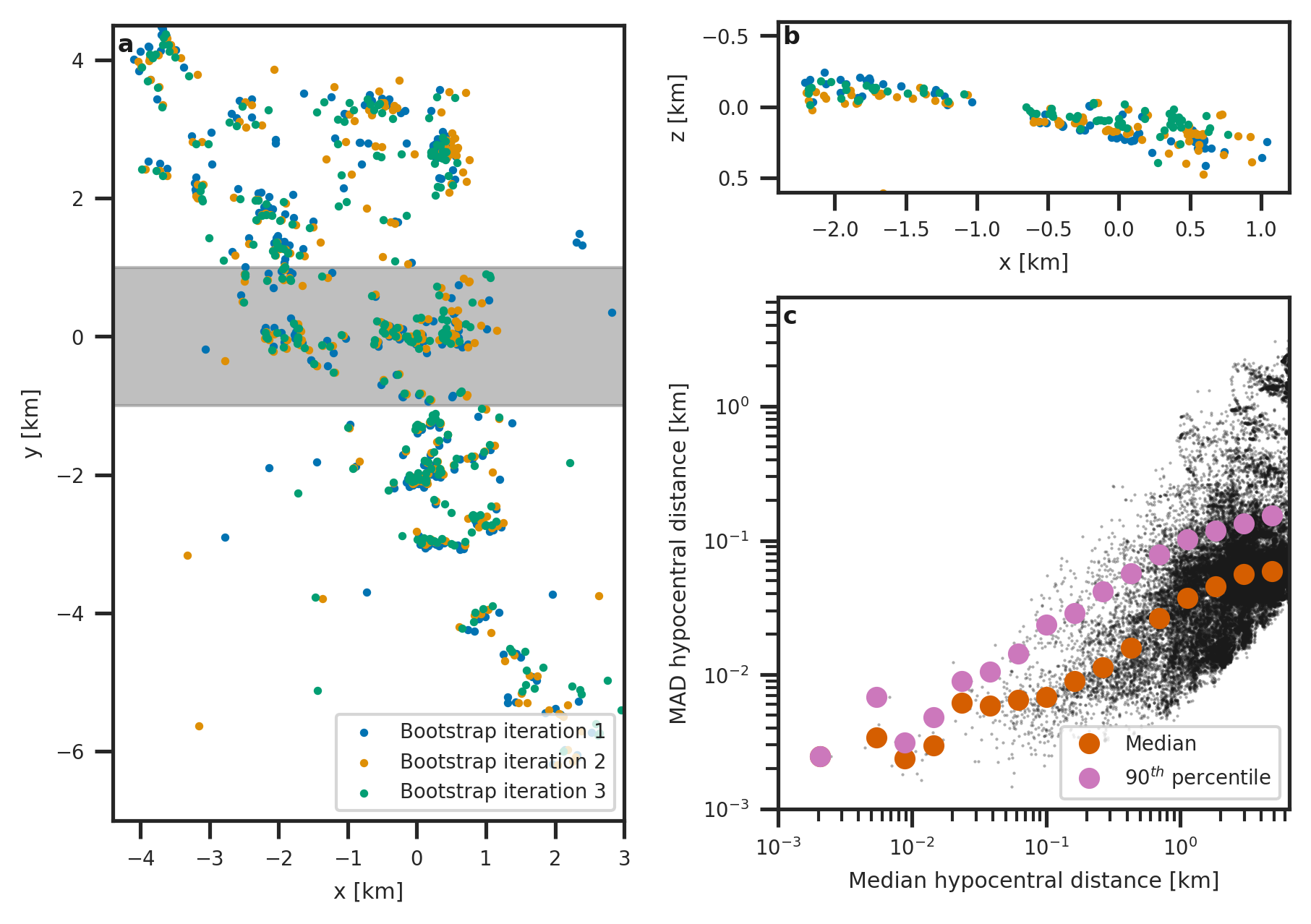}
\caption{Relocation bootstrap analysis with GrowClust3D and the bootstrapping implementation of GrowClust3D. The analysis was conducted on all events located between 72\degree W and 71\degree W and between 28\degree S and 26.9\degree S. We show the initiation cluster of the 2023 SSE sequence around 71.3\degree W 27.3\degree S. All events have been transformed to a local coordinate system, the absolute location offset between the individual bootstrap iterations has been removed. \textbf{a} Epicenter distribution in the first 3 bootstrap iterations. \textbf{b} Cross section through the cluster. The events shown are highlighted in grey in subplot a. \textbf{c} Statistical results from 1000 bootstrap iterations. Each black dot represents one event pair with their median hypocentral distance and the median absolute deviation of the distance as an uncertainty measure. Red and purple dots show the moving median and $90^{th}$ percentile of the uncertainties.}
\label{fig:relocation_bootstrap}
\end{figure}

\begin{figure}[ht!]
\centering
\includegraphics[width=\textwidth]{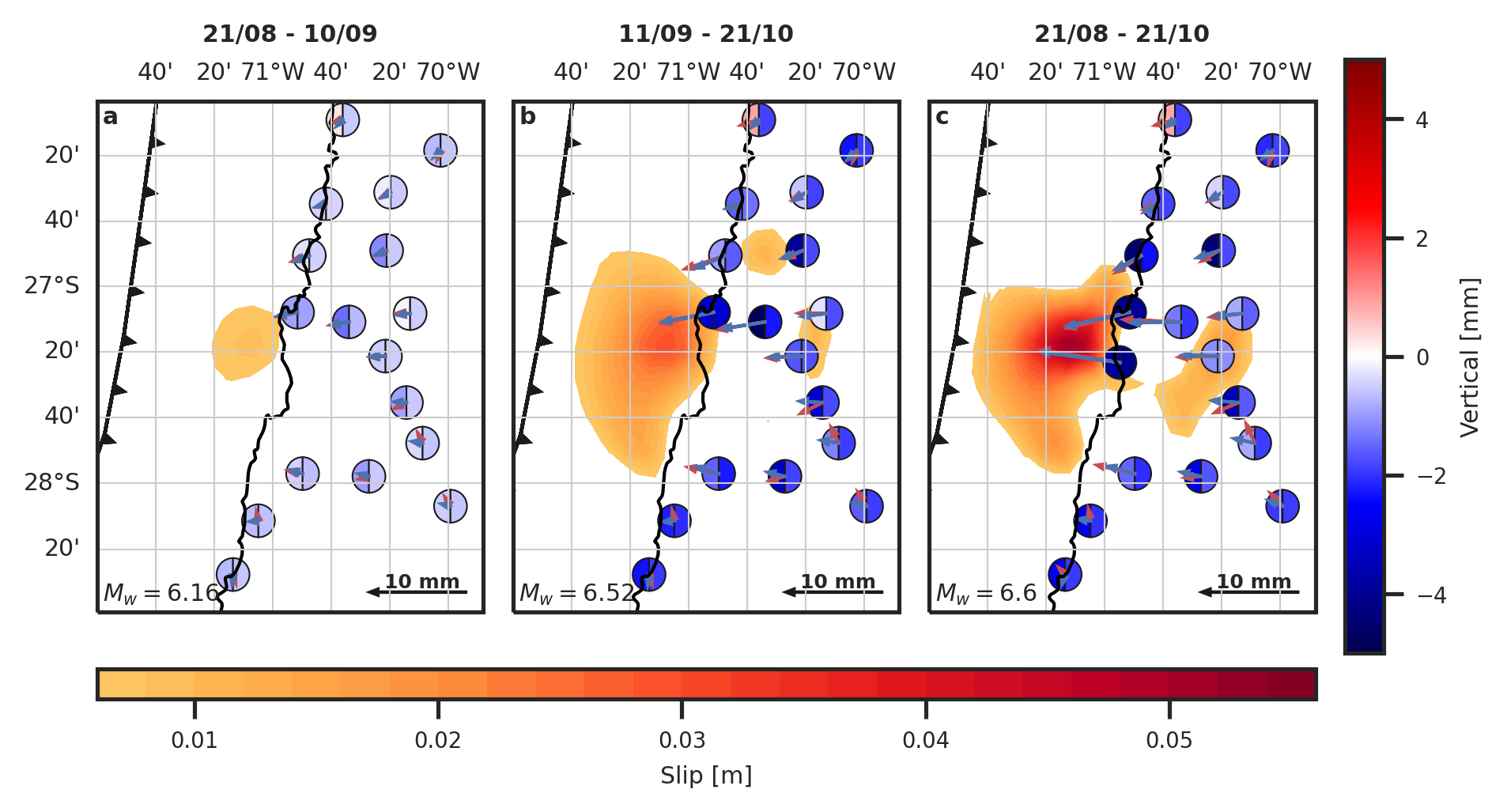}
\caption{Snapshots of the slip for 3 periods. The arrows provide a comparison between observed (blue arrows) and predicted (red arrows) slip in three phases. Circles denote vertical displacement, with the left half representing the observed and the right half the predicted displacement.}
\label{fig:gnss_match_phase}
\end{figure}

\begin{figure}[ht!]
\centering
\includegraphics[width=\textwidth]{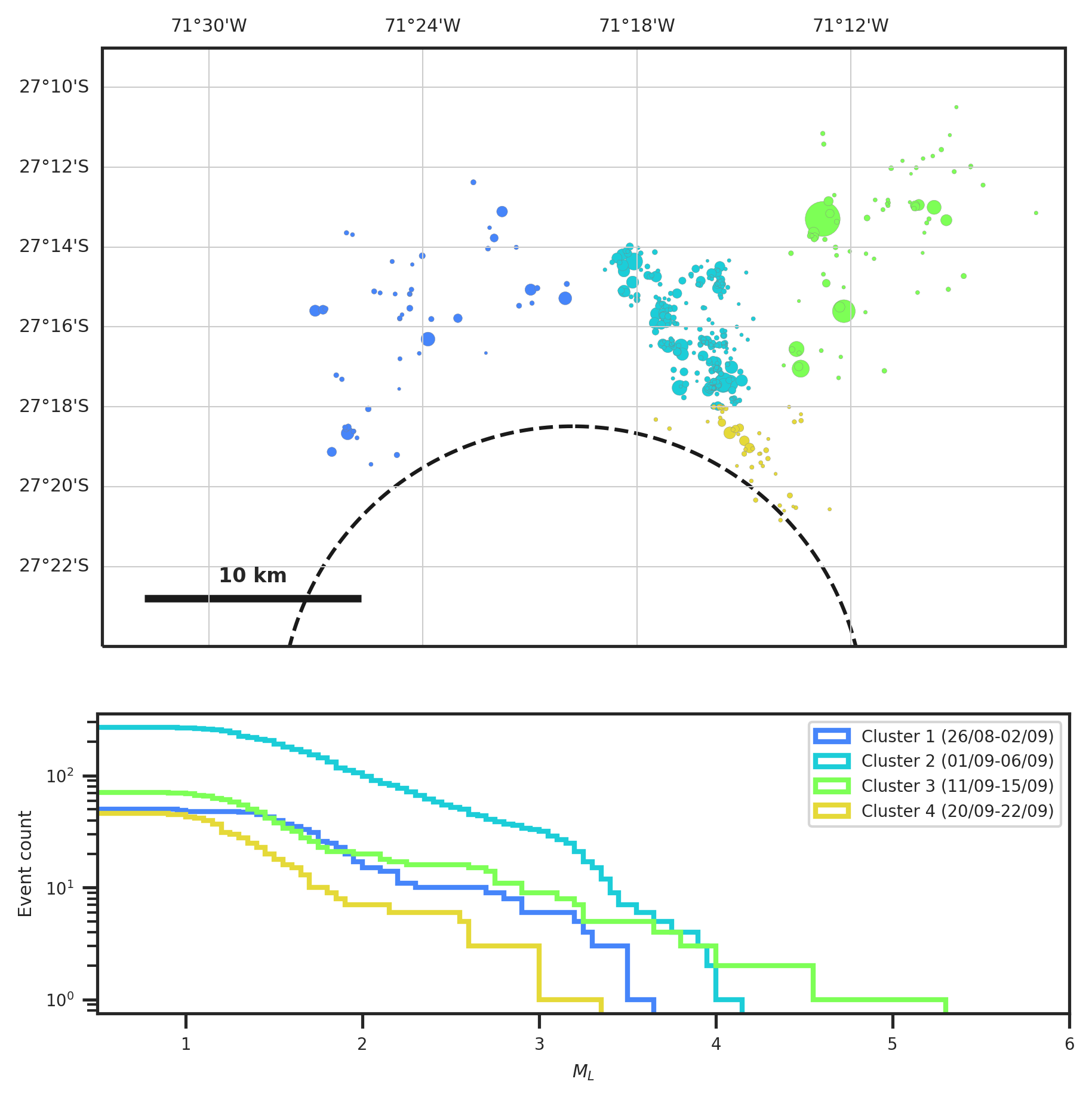}
\caption{The onset sequence grouped into individual clusters with locations and magnitude distribution. The blue cluster occurs first (26/08-02/09), followed by the cyan (01/09-06/09), the green (11/09-15/09), and the yellow last (20/09-22/09). The bottom plot shows the inverse cumulative magnitude distribution. The black dashed circle indicates the approximate position of the subducted seamount inferred from bathymetry and gravity data.}
\label{fig:onset_clusters}
\end{figure}

\begin{figure}[ht!]
\centering
\includegraphics[width=\textwidth]{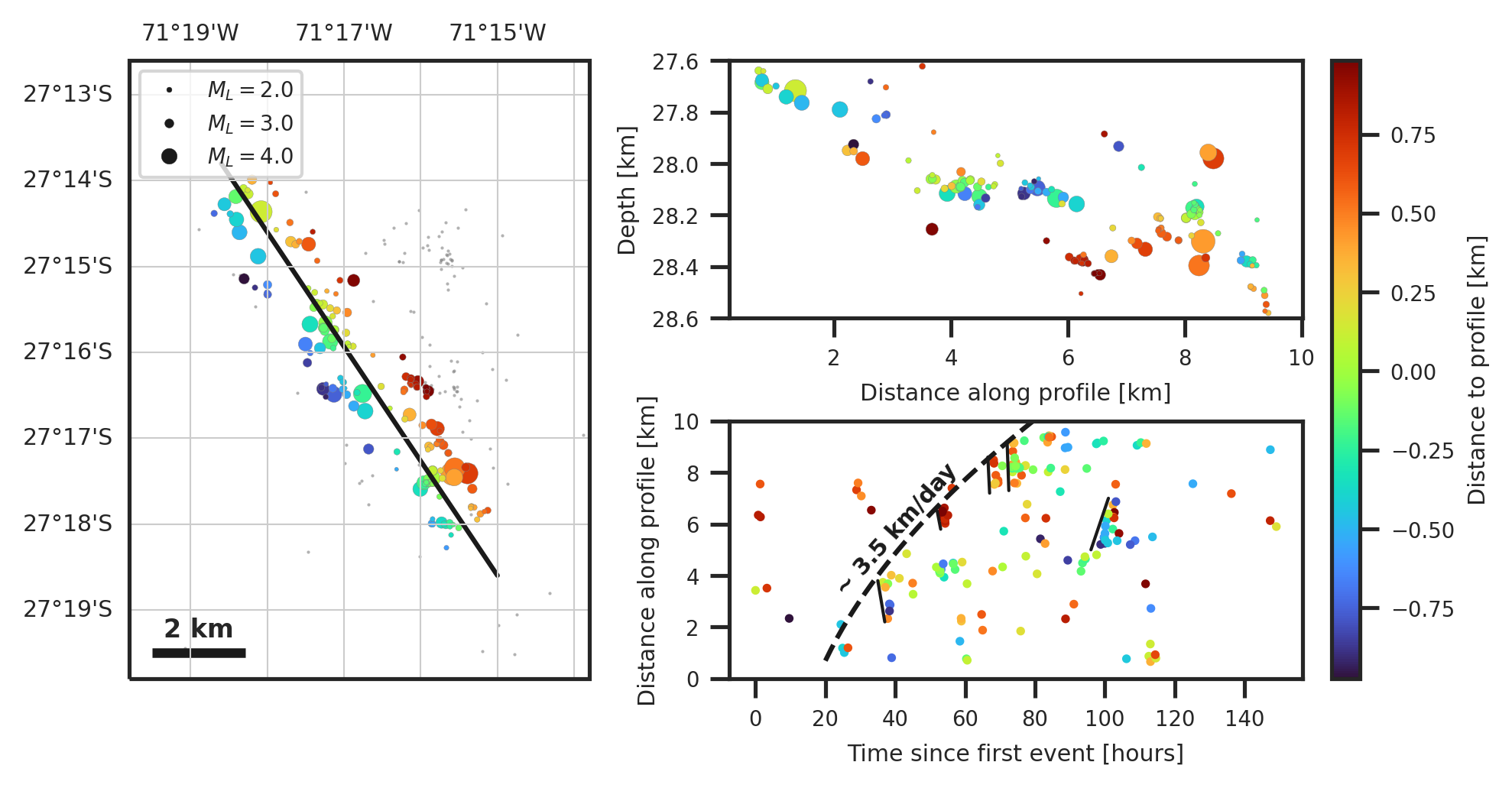}
\caption{Zoom in on the highly active cluster during the onset sequence. The black line in the map plot shows the transect used in the two other plots. The transect is 2~km wide and centered around the line. All events within the transect are shown in colour on the map. We highlight the slow (dashed line) and fast (narrow, solid lines) migrations. For the fast migrations, we are not able to determine a reliable migration velocity estimate, but it is in excess of 15 km/day.}
\label{fig:onset_detail_v2}
\end{figure}

\begin{figure}[ht!]
\centering
\includegraphics[width=0.7\textwidth]{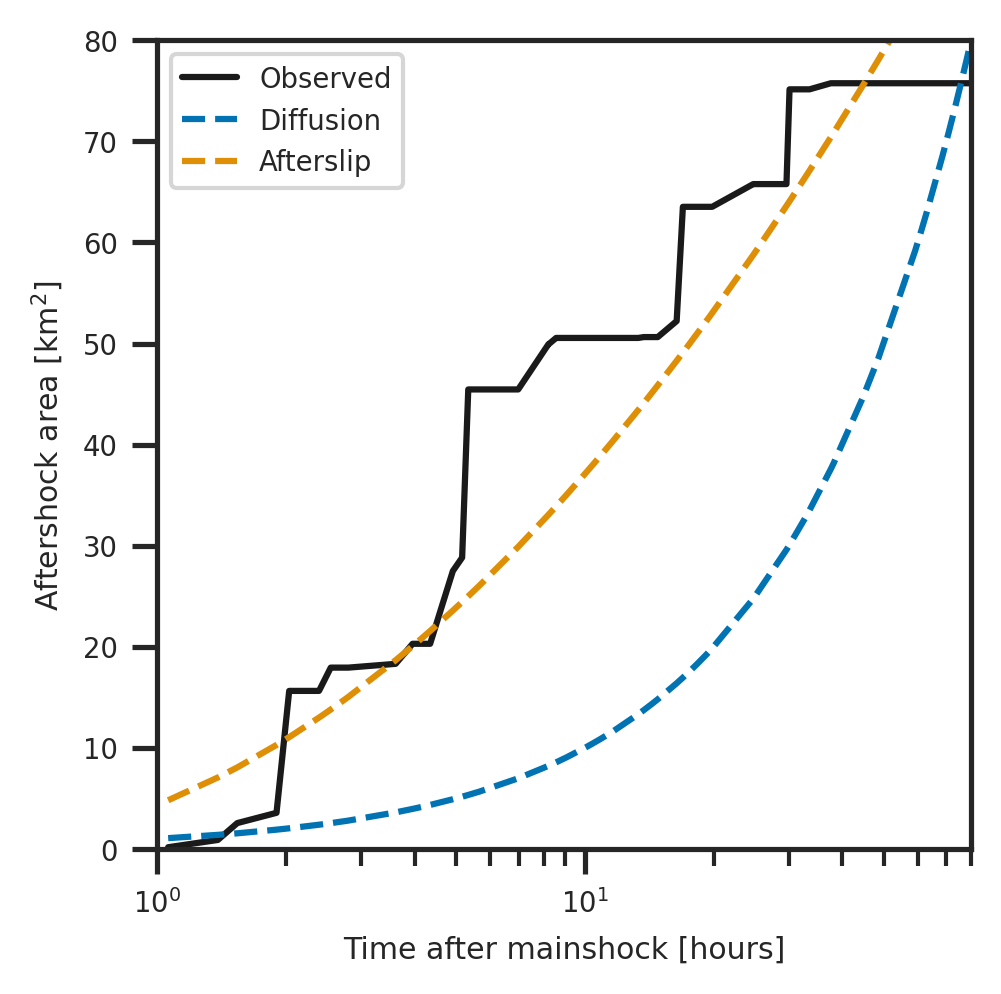}
\caption{Area of the aftershocks for the $M_L=5.3$ event on September 11th. The area is estimated using the convex hull of all aftershocks so far. For comparison, prototypical expansion curves for afterslip (4~km per decade) \cite{perfettiniModelAftershockMigration2018} for a diffusion process (280~m$^2$/s) \cite{shapiroTriggeringSeismicityPorepressure2003} have been added.}
\label{fig:onset_cluster3}
\end{figure}

\begin{figure}[ht!]
\centering
\includegraphics[width=\textwidth]{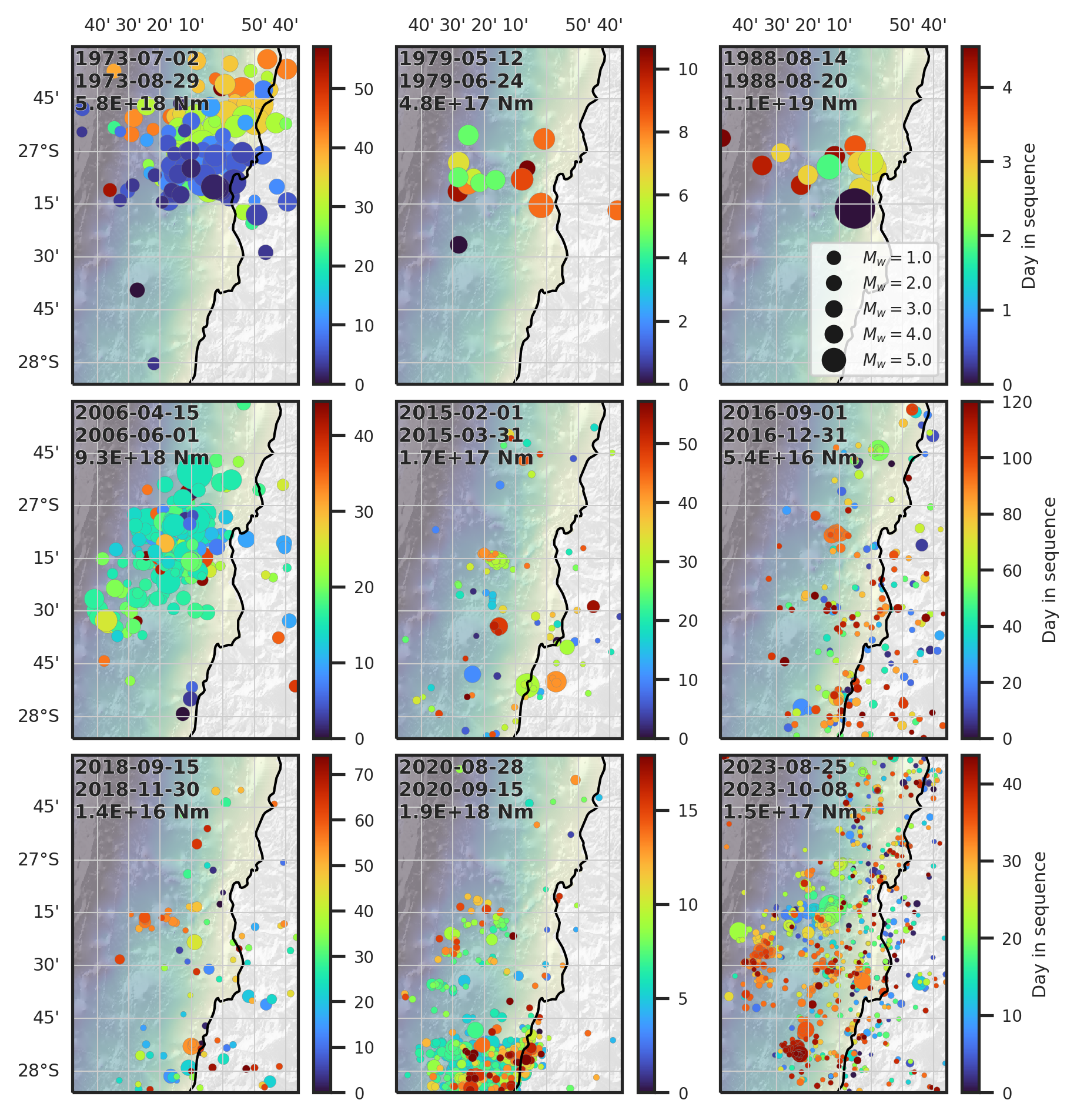}
\caption{Historic records of seismic swarms and migrations in aftershock sequences. The 1973, 1979 and 1988 sequences use the USGS catalog (covering 1964 - 2024), the 2006 sequence the CSN catalog (2000 - 2024), the 2015 to 2020 sequences uses the permanent station catalog from \citeA{munchmeyer2024chile_eqs} (2014 - 2024), and the 2023 sequence the temporary station catalog from \citeA{munchmeyer2024chile_eqs} (11/2020 - 02/2024). The total seismic moment of all visualised events per sequence is indicated in the top left corner. A zoom in on the 2015 to 2023 swarms is available in Figure~4.}
\label{fig:historic_swarms}
\end{figure}

\end{document}